\documentstyle[11pt]{article}

\renewcommand{\theequation}{\arabic{equation}}
\newcommand{\be}{\begin{equation}}
\newcommand{\ee}{\end{equation}}
\newcommand{\bea}{\begin{array}}
\newcommand{\ea}{\end{array}}
\newcommand{\beqa}{\begin{eqnarray}}
\newcommand{\eeqa}{\end{eqnarray}}
\newcommand{\bean}{\begin{eqnarray*}}
\newcommand{\eean}{\end{eqnarray*}}
\newcommand{\eqn}[1]{(\ref{#1})}

\newcommand{\del}{\partial}
\newcommand{\nn}{\nonumber}

\def\up#1{\leavevmode \raise.16ex\hbox{#1}}

\def\sqr#1#2{{\vcenter{\vbox{\hrule height.#2pt
	\hbox{\vrule width.#2pt height#1pt \kern#1pt
	  \vrule width.#2pt}
	\hrule height.#2pt}}}}

\newcommand{\journal}[4]{{\sl #1 }{\bf #2} \up(19#3\up) #4}

\newcommand{\gapproxeq}{\lower .7ex\hbox{$\;\stackrel{\textstyle >}{\sim}\;$}}
\newcommand{\lapproxeq}{\lower .7ex\hbox{$\;\stackrel{\textstyle <}{\sim}\;$}}

\renewcommand{\theequation}{\thesection.\arabic{equation}}

\newcommand{\appendice}
{
\setcounter{equation}{0}
\renewcommand{\theequation}{A.\arabic{equation}}
}

\def\thebibliography#1{{\bf REFERENCES\markboth
 {REFERENCES}{REFERENCES}}\list
 {[\arabic{enumi}]}{\settowidth\labelwidth{[#1]}\leftmargin\labelwidth
 \advance\leftmargin\labelsep
 \usecounter{enumi}}
 \def\newblock{\hskip .11em plus .33em minus -.07em}
 \sloppy
 \sfcode`\.=1000\relax}

\begin{document}

\title{Deformed Chern--Simons Theories}
\author{G. Bimonte$^{a}$, R. Musto$^{a}$, A. Stern$^{b}$ and
P. Vitale $^{a}$}

\maketitle

\begin{center}
{\it a)  Dipartimento di Scienze Fisiche, Universit\`a di Napoli,\\
Mostra d'Oltremare, Pad.19, I-80125, Napoli, Italy; \\
INFN, Sezione di Napoli, Napoli, ITALY.\\
\small e-mail: \tt bimonte,musto,vitale@napoli.infn.it } \\
{\it b) Department of Physics, University of Alabama,\\
Tuscaloosa, AL 35487, USA.\\
\small e-mail: \tt astern@ua1vm.ua.edu }\\
\end{center}

\begin{abstract}
We construct a Chern-Simons action for $q$--deformed gauge theory which
is a simple and straightforward generalization of the usual one.
Space-time continues to be an ordinary (commuting) manifold,
while
the gauge potentials and the field strengths become $q$--commuting fields.
Our approach, which is explicitly carried out for the case of `minimal'
deformations, has the advantage of leading naturally to
 a consistent Hamiltonian structure that has essentially
 all of the features of the   undeformed case.  For example,
 using the new Poisson brackets, the constraints form a
closed algebra and generate $q$--deformed gauge transformations.
\end{abstract}

\hfill

\small DSF 16/97$~~~~~~~~~~~~$
UAHEP 975$~~~~~~~~~~~~$

\newpage

\section*{Introduction}
\setcounter{equation}{0}
Chern-Simons (CS)  theory has played a unique r\^ole in unifying
different, previously unrelated, physical problems and mathematical ideas
in $1+1$ and $2+1$ dimensions.
 CS theory gives a natural description of anyonic
excitations of condensed matter systems \cite{stat} constrained to two space
dimensions, and it
provides an important key to understanding and classifying the
intriguing physics of quantum Hall fluids \cite{qhe}.
Also, unique to three dimensions is the  possibility of expressing
quantum gravity as a pure CS theory written in terms of dreibeins and
spin-connections transforming locally under the Poincar\'e group
\cite{wit1,at}.

Here  our motivation is concerned with the r\^ole the CS action
provides in describing mathematical and physical
problems usually formulated in two dimensions, which
was first discussed by E.
Witten in ref. \cite{wit2}. By showing that quantum CS
theory provides a framework for understanding Jones polynomials of
knots in three
dimensions, he also shed new light on conformal theories in $1+1$
dimensions \cite{moore}. In making
the correspondence between the two and three dimensional
theories one can  consider CS theory on a manifold with a boundary, such
as a disk $D\times {\bf R}^1$, ${\bf R}^1$ being the time line.
Then all
states in the bulk can be gauged away and one is left with a family of
conformal states, called edge states,  on the boundary.
An extensive literature has been devoted
to the study of edge states \cite{edge}.
In particular, the edge states of quantum Hall
devices at fractional fillings, which have been  analyzed from
different theoretical points of view \cite{wcw}, provide a unique experimental
laboratory for probing non-fermi liquids.
Here we recall  two
theoretical aspects of edge states. First, they  define
a Kac-Moody algebra
which is easily derived from the Hamiltonian formulation of the theory.
Second,
the essential tool for building affine algebras starting from conformal
field theory \cite{god}, namely the Fubini-Veneziano vertex
operator \cite{fubini},
has a natural realization in terms of the Wilson line for
CS theory \cite{bal2}.

While this picture relating $2+1$ and $1+1$ physics, which we briefly
sketched, has  been greatly developed and clarified, new ideas
have emerged in $1+1$ dimensions whose corresponding r\^ole (if any)
in $2+1$ dimensions remains
unclear.  In this regard, $q$-deformed
affine Lie algebras associated with quantum groups \cite{djfm}
have been formulated for the entire non-exceptional
series and a
construction in terms of anyonic $q$-deformed oscillators has been given, at
least for the unitary and symplectic cases
\cite{scia}.   Furthermore, $q$-deformed affine Lie
algebras, enter in different aspects of $1+1$ integrable
models \cite{inte,RSV}. In particular,
they appear as the minimal symmetry needed to
determine the $\cal S$-matrix up to an overall scalar factor.
Such algebraic structures naturally arise
\cite{felt} when conformal models
are driven off criticality by an appropriate perturbation preserving the
integrability \cite{zamol}
of the theory.  Here  $q$ is related
to the value of the coupling appearing
in the generalized vertex operators.

Without expanding further on
different aspects of these developments,  it seems
fair to say that  a complete  picture is still lacking, and therefore it
may be useful to look at them from the new
 perspective of a $2+1$ dimensional theory.
It is with this in mind, that we  construct a CS action
for $q$-deformed gauge theory as a simple and straightforward generalization
of the undeformed case.

We note that in  \cite{cin} deformed Chern characters are
constructed for $SU_q(2)$, based on a
different differential calculus on quantum groups  \cite{wata}.
 The resulting CS action differs from ours,
in many respects.  Theirs requires a trilinear metric,
and extra conditions placed on the quantum group metric, neither of
which are  needed in our approach.   Furthermore,
our approach leads naturally to
 a consistent Hamiltonian structure that has all the features present
 in the ordinary case.   Analogous to the undeformed case, first class
 constraints appear in the formalism which
generate the $q$-deformed gauge transformations  and
form a closed algebra.
Our construction of  the $q$-deformed CS
action, with its natural Hamiltonian structure,
 paves the way to an analysis of  $q$-CS edge
states, which we will discuss in a separate paper \cite{prep1}.
It is also very simple in
our framework to discuss $q$-deformed general relativity in $2+1$
dimensions \cite{prep2}, where we desire a departure from the
undeformed theory ($q=1$) at Planck scale curvatures.

The paper is organized as follows. In Section $1$, we briefly review the
structure of $q$-deformed gauge theories mainly along the line of ref.
\cite{cast1}.
In Section $2$, we show that a $q$--CS action can be defined
in the case of a {\it minimal} deformation.
The requirement that the theory be minimally
deformed is sufficient (and perhaps necessary, as well) to ensure the gauge
invariance of the CS action.
(Minimality is also needed for the closure of gauge transformations.)
In Section $3$, it is shown that the solutions of the
equations of motion are, as expected, flat connections and a number of
useful properties are exhibited, among them the relation with $q$-deformed
Pontryagin densities.
Finally, the last Section is devoted to the
construction of
a consistent Hamiltonian formalism. By introducing deformed Poisson brackets
for the components of the connection we show that the field strengths,
smeared
with Lie algebra valued functions generate gauge transformations and lead to
a closed algebra. Details of calculations are given in the appendix.
 We close with brief final remarks.

\section{$q$--Deformed Gauge Theories}
\setcounter{equation}{0}
In this Section we briefly review the mathematical setting which is
needed to introduce gauge field theories whose infinitesimal gauge
symmetry is associated with a quantum Lie algebra.

Recently there have been various proposals for such $q$--deformed
theories \cite{cast1}-\cite{bsv}.
In some of them \cite{brez2} the structure of space--time is made
noncommutative, which is
especially relevant for applications to gravity. In other
theories, \cite{cast1,brez1,isa},  the structure
of space--time stays commutative and a bicovariant differential
calculus \cite{woro}-\cite{ac} is needed in order
to define the quantum Lie algebra. In \cite{sud} an $SU_q(2)$ gauge
theory is proposed that is based on a definition of a quantum Lie algebra
which doesn't need a bicovariant differential calculus, while \cite{bsv}
contains a proposal of an $SU_q(2)$ gauge theory on the lattice.
We will follow in the paper the approach of Castellani, because it seems
to us closer to usual gauge theory and differential
calculus on classical Lie groups.

Let us first recall the definition of
  a quantum Lie algebra and its connection to
differential calculus on quantum groups, as described in \cite{ac}.
Starting from the definition
of a quantum group $G_q$ as the noncommutative algebra of functions
on the Lie group $G$, $G_q\equiv Fun_q(G)$, a bimodule of left
(right) invariant forms for $G_q$ is constructed, in the same way as
the bimodule of left (right) invariant forms is constructed for
classical Lie groups. Such a bimodule inherits the noncommutative
nature of the product in $Fun_q(G)$
\be
R^{ab}_{ef} \, M^e_c M^f_d = M^b_f  M^a_e \,
R^{ef}_{cd}~,
\label{rtt}
\ee
($M$ is an element of $G$ in its defining representation)
so that the usual definition of exterior product for one--forms
, $ \theta^i \wedge \theta^j =  \theta^i \otimes \theta^j -
\theta^j \otimes\theta^i$, is replaced on $q$--groups by
\be
\theta^i \wedge \theta^j = \theta^i \otimes \theta^j -
\Lambda^{ij}_{kl} \, \theta^k \otimes \theta^l
\label{formprod}
\ee
where $\Lambda$ is the braiding matrix.
Following the analogy with the differential calculus on classical Lie
groups, the algebra of left invariant vector fields, which  is dual to
the algebra of left invariant one-forms, can be obtained,\footnote{In
the same way we can introduce right invariant objects. The
differential calculus which is at the basis of the deformed gauge
theory we are going to consider is bicovariant; bicovariance requires
 that left and right actions of the $q$--group on the bimodule
commute.}
with $q$--commutation relations
\be
T_i T_j - \Lambda_{ij}^{kl} \, T_k T_l \equiv [T_i, T_j]=C_{ij}^k T_k ~.
\label{qcom}
\ee
It is this algebra which is called a quantum Lie algebra. In the limit
$q\rightarrow 1$, $\Lambda^{kl}_{ij} \rightarrow \delta^k_j
\delta^l_i$ and $T_i$ become the generators of the classical Lie
algebra.  $C_{ij}^k$ are $q$-structure constants, which in general are
not antisymmetric in the lower two indices except in the limit
$q\rightarrow 1$. (In principle, the $q$--Lie
algebra might be bigger than the classical Lie algebra, its dimension
being equal to the dimension of the bimodule.)

In order to define a bicovariant calculus the braiding matrix $\Lambda$ and
the structure constants have to satisfy the following relations
\cite{ac}:
\beqa
\Lambda^{ij}_{kl}\Lambda^{lm}_{sp} \Lambda^{ks}_{qu} =
\Lambda^{jm}_{kl}\Lambda^{ik}_{qs} \Lambda^{sl}_{up}~~~
({\mbox{Yang~Baxter~
equation}}) \label{yb} \\
C^r_{mi} C^n_{rj}  -\Lambda^{kl}_{ij} C^r_{mk} C^n_{rl} =
C_{ij}^k C_{mk}^n~~~({\mbox{$q$-Jacobi}}) \label{qjac} \\
\Lambda^{ir}_{mk}\Lambda^{ks}_{nl} C_{rs}^j = \Lambda^{ij}_{kl} C_{mn}^k
~,
\label{2d} \\
\Lambda^{jq}_{ri}\Lambda^{si}_{kl} C_{ps}^r + \Lambda^{jq}_{pi} C_{kl}^i=
 C_{is}^j \Lambda^{sq}_{rl}\Lambda^{ir}_{pk} + C_{rl}^q\Lambda^{jr}_{pk}~.
\eeqa
The first condition is the quantum Yang Baxter equation; the second
is the Jacobi identity for the algebra \eqn{qcom}, while the
last equations are trivial in the limit $q\rightarrow 1$.

Following \cite{cast1}, the gauge potential is assumed to be a
$q$--Lie algebra valued one--form
$A \equiv A^i_\mu T_i dx^\mu$ (we will often write
$A^i$ to mean the one--form $A^i= A^i_\mu dx^\mu$).
In this approach the deformation occurs solely in the fiber and
thus the $A^i_{\mu}$ are taken to be
$q$--fields subject to nontrivial commutation relations.
Space--time, instead, remains an ordinary manifold so that $d x^{\mu}$ are
ordinary space--time differentials commuting with $A^i_{\mu}$.
The exterior product of one--forms on the space--time  manifold is deformed
in the
same way as the exterior product of invariant forms on the group manifold
\eqn{formprod} and, for general groups, one has
\cite{mont}:
\be
A^i \wedge A^j = - Z^{ij}_{kl} \, A^{k} \wedge A^l~;  \label{com}
\ee
where $Z$ is a matrix of ordinary $c-$numbers which depends on the
group. The undeformed
case obviously corresponds to the choice $Z^{ij}_{kl}=\delta^i_l
\delta^j_k$ for any group.  It is determined in general by insisting
that  $Z^{ij}_{kl}+\delta^i_k
\delta^j_l$ is proportional to a projection operator, so that there are
no further restrictions on $A^i \wedge A^j$.
In the rest of this Section and the beginning of the next
we shall consider the general deformation
of $U(n)$.  [We shall later specialize to a particular type of
deformation known as {\it minimal}.]
For the general deformation
of $U(n)$,  the matrix $Z$ has a simple expression
in terms of the braiding matrix $\Lambda$
\cite{cast1}:
\be
A^i \wedge A^j = - {1\over r^2 + r^{-2}} (\Lambda + 
\Lambda^{-1} )^{ij}_{kl}\,A^{k} \wedge A^l~,  \label{qaa}
\ee
where $r$ is a deformation parameter; we are assuming multiparametric
deformations as considered in \cite{sch}. The braiding matrix $\Lambda$
will depend in general on a set of parameters $q_{i}$ and on $r$. The number of
independent parameters depends on the group
(to make contact with the $U_q(2)$ gauge theory of \cite{cast1} one has to
remember that in this case there is only one parameter).
The commutators of the $q$--fields
$A^i_{\mu}$ follow from eq.(\ref{qaa}), after one
factorizes the coefficients $d x^{\mu} \wedge dx ^{\nu}$:
\be
A^i_{[ \mu} A^j_{\nu ]} = - {1\over r^2 + r^{-2}}
(\Lambda + \Lambda^{-1} )^{ij}_{kl}\,
A^{k}_{[\mu} A^l_{\nu ]}~.
\ee
From now on we will omit the symbol $\wedge$ for product of forms.
The deformed gauge transformations are assumed to be of the usual form
\be
\delta_{\epsilon} A= -d\epsilon -A\epsilon + \epsilon A \label{var1}
\ee
where $\epsilon \equiv \epsilon^i T_i$. The gauge parameters
$\epsilon^i$ are now $q$--numbers and are assumed to have the following
commutation rules with the gauge fields:
\be
\epsilon^i A^j = \Lambda^{ij}_{mn} A^m \epsilon^n~. \label{com1}
\ee
The commutation relations for $A^i$ with $d\epsilon^j$ and $dA^i$
with $\epsilon^j$ can be obtained by taking the exterior derivative
of the above equation and imposing that the terms containing $dA^i$
and $\epsilon^j$ cancel separately.
The field strength is defined in the usual way
\be
F \equiv \frac{1}{2} F_{\mu \nu} dx^{\mu} dx^{\nu}
 = d A + A ^2~,   \label{fistrength}
\ee
where $A^2 = A^i A^j T_i T_j$.
$F$ is an element of the deformed Lie-algebra \cite{cast1} and
under a gauge transformation \eqn{var1} it transforms as:
\be
\delta_{\epsilon} F = \epsilon F - F \epsilon~.\label{var2}
\ee

In \cite{cast1} it is shown that the
$q$-Lagrangian ${\cal L}= <F_{\mu \nu},F^{\mu \nu}>_q$ is
invariant under the transformation \eqn{var1},
if the $q$-deformed scalar
product on the quantum Lie algebra
$<\cdot,\cdot>_q$ obeys the following invariance
condition
\be
\Lambda_{rs}^{nj} <[T_m, T_n], T_j>_q + <T_m, [T_r, T_s]>_q = 0
\label{scalinv},
\ee
which generalizes the invariance property of the Killing metric
on a Lie algebra. Upon introducing the
matrix $g_{ij}=<T_i,T_j>_q$,
the above equation can also be written as
\be
\Lambda_{rs}^{nj} C_{mn}^i g_{ij} + C_{rs}^j g_{mj} = 0~.
\ee
It is interesting to notice that the deformed metric $g_{ij}$
is not symmetric, in general: $g_{ij} \neq g_{ji}$.
We will show that this invariance condition
is necessary, but not sufficient to construct a CS Lagrangian
with $q$--symmetry.   In Section 3, we will make an additional
assumption on the theory whereby a certain combination of $g_{ij}$ and
$\Lambda_{rs}^{nj} $ is nondegenerate.

\section{The Chern--Simons Lagrangian density}
\setcounter{equation}{0}
In this Section we will try to construct a
deformed CS Lagrangian density in the framework of the deformed gauge
theories sketched in the previous Section.
So, we search for a Lagrangian density ${\cal L}_{CS}$ such that:

\noindent
1) ${\cal L}_{CS}$ is a three-form that  changes by a total derivative
 under an infinitesimal gauge transformation (\ref{var1});

\noindent
2) $d{\cal L}_{CS}=<F,F>_q$;

\noindent
3) The equations of motion for the $q$--CS action 
are the zero curvature conditions $F=0$.

Inspired by the classical formula, we make the following
ansatz for the deformed CS Lagrangian density
\be
{\cal L}_{CS}= <dA + \beta A^2, A>_q   \label{lag}
\ee
where $\beta$ is a factor to be determined
(we shall omit writing, from now on, the subscript $q$).

We start by checking 1) whether, for any choice of $\beta$, the variation
of eq.(\ref{lag}) under an infinitesimal $q$--gauge transformation
(\ref{var1}) is a total derivative. When one performs such a transformation
in eq.(\ref{lag}) two types of terms arise, those containing $\epsilon$
and those containing $d \epsilon$. We shall collect them separately
and accordingly split $\delta_{\epsilon} {\cal L}_{CS}$ as:
\be
\delta_{\epsilon} {\cal L}_{CS}=
\delta {\cal L}_1
+ \delta {\cal L}_2,~ \label{var3}
\ee   
with the $\epsilon$ dependent terms in $\delta {\cal L}_1$ and
the $d\epsilon$ dependent terms in $\delta {\cal L}_2$.
An easy computation gives:
\beqa
\delta {\cal L}_1 &=& <dA, \epsilon A-A\epsilon> + <- dA \epsilon +\epsilon
dA, A> + \nn\\
&+& \beta <\epsilon A^2 - A^2 \epsilon, A> + \beta <A^2,
-A\epsilon +\epsilon A>~, \label{vlag1}
\eeqa
and
\be
\delta {\cal L}_2 = -<dA, d\epsilon> + (1-\beta)<d\epsilon A + A d\epsilon, A>
- \beta < A^2, d\epsilon> ~.
\label{vlag2}
\ee
$\delta {\cal L}_1$ vanishes, by virtue of the invariance
condition for the scalar product \eqn{scalinv}. In fact, using the
commutation rules eqs.(\ref{qcom}), (\ref{com1})
the sum of the first two terms of the right hand side
of eq.(\ref{vlag1}) can be rewritten as:
$$
<dA, \epsilon A-A\epsilon> + <- dA \epsilon +\epsilon dA, A>  =
$$
\be
=-dA^i A^m \epsilon^n \{ <T_i, [T_m, T_n]> + \Lambda^{jk}_{mn}
<[T_i, T_j], T_k> \}
\ee
which is zero because of eq.(\ref{scalinv}).
As for the sum of the third and fourth terms in the right
hand side of eq.(\ref{vlag1})
it can be shown to vanish using the following commutation relations,
proven in \cite{cast1}:
\be
\epsilon^i (A^2)^j =  \Lambda^{ij}_{mn} (A^2)^m \epsilon^n. \label{com2}
\label{com3}
\ee
Using this, we get
$$
<\epsilon A^2 - A^2 \epsilon, A> +  <A^2,
-A\epsilon +\epsilon A> =
$$
\be
=- (A^2)^i A^j \epsilon^k \{\Lambda^{tm}_{jk} < [T_i, T_t], T_m >
+ \, < T_i, [T_j, T_k] >\} =0.
\ee
We are thus left with
$\delta {\cal L}_2$. From eq.(\ref{vlag2}) we see that, among
the three terms appearing in it, only the first
one is a total derivative; in order to see whether the remaining two
vanish we first rewrite them in a convenient way.
Using the commutation properties
of $d\epsilon$ with $A^i$, the second term can
be rewritten as
\beqa
< d\epsilon A + A d \epsilon, A > & = &
- A^i A^j d\epsilon^k \Lambda^{qm}_{jk} < [T_i, T_q], T_m >=\\
&=& A^i A^j d\epsilon^k < T_i, [T_j, T_k] >~.  \label{v1}
\eeqa
Using instead the explicit expression  for $A^2$ \cite{cast1}
\be
(A^2)^i = {1\over (2 + r^2 + r^{-2} ) } A^j A^k [ C_{jk}^i -
(\Lambda^{-1})^{lm}_{jk} C_{lm}^i] \label{a2}
\ee
we get for the third term of eq.(\ref{vlag2})
\be
<  A^2, d\epsilon > =
{1 \over (2+ r^2 + r^{-2})} A^i A^j d\epsilon^k
\{<[T_i, T_j], T_k> - (\Lambda^{-1})^{ml}_{ij} < [T_m, T_l], T_k > \}.
\label{v2}
\ee
Summing eq.(\ref{v1}) to eq.(\ref{v2}) we get
$$
\delta {\cal L}_{CS} =- <dA, d \epsilon> +
$$
\be
 + A^i A^j d\epsilon^k \left\{
(1-\beta) < T_i, [T_j, T_k]  >
- {\beta\over (2+ r^2 + r^{-2})}
(<[T_i, T_j], T_k> - (\Lambda^{-1})^{ml}_{ij} < [T_m, T_l], T_k > )
\right\}. \nn\\
\label{lv}
\ee
In order for $\delta_{\epsilon}{\cal L}_{CS}$ to be a total derivative,
the expression between the curly brackets must vanish.     However,
it does not vanish in general, as we have checked that it is
different from zero for the case of $U_q(2)$,
using the explicit formulae in \cite{cast1}.  On the other hand,
it can be proven to be zero if we make a further
assumption on the theory; namely, that
 the deformation be $\it minimal$, which means that
\be
\Lambda^{ij}_{kl} \, \Lambda^{kl}_{mn}  =\delta^i_m \, \delta^j_n
. \label{min}
\ee
In this case, the anticommutation relations eq.(\ref{com}) can be derived
directly from eq.(\ref{formprod}) [setting $\theta^i=A^i$]
by multiplying both sides by
$\Lambda^{mn}_{ij}$ (and summing over $i$ and $j$), so that we have
\be
A^m A^n = - \Lambda^{mn}_{ij} A^i A^j~,~~~\mbox{for~minimal~deformations}
.\label{aa}
\ee
Consistency with eq.(\ref{qaa}) then implies $r^2=1$.
The simple commutation relations above allow us to
write $A^2$ in a form analogous to the undeformed case:
\be
A^2 = {1\over 2} A^i A^j [T_i, T_j]~.\label{newa2}
\ee
In the minimal case, one can also prove a number of further simplifying
relations.  For example, by multiplying eq.(\ref{qcom})
by $\Lambda$ on the right, one finds that the $q$--structure
constants $C^k_{ij}$ are $\Lambda$-antisymmetric \cite{cast1}
\be
C_{ij}^k= - \Lambda_{ij}^{rs} C_{rs}^k  , ~~~~
C_{ij}^k= - (\Lambda^{-1})_{ij}^{rs} C_{rs}^k \label{skewsym}.
\ee
Moreover, multiplying eq.(\ref{scalinv}) by
$\Lambda^{-1}$ and using eq.(\ref{skewsym}), one gets the following
usual expression for the invariance condition of the inner product:
\be
< [T_i, T_j], T_k > =   < T_i, [T_j, T_k] > ~.
\label{scalinv2}
\ee
Using eq.\eqn{skewsym} and eq. \eqn{scalinv2} into eq.\eqn{lv}
one finds:
\be
\delta {\cal L}_{CS} =-<dA,d \epsilon>\, + \, \left(2-3\beta \right)
<A^2, d\epsilon>~.
\ee
The second term in the r.h.s. is zero if $\beta=2/3$ and this is the
value we choose. With this choice, our deformed
CS Lagrangian density has the same expression as the undeformed one; this
resemblance is only formal, as the new Lagrangian is written in terms of
a deformed scalar product, and the gauge fields are noncommuting.

Even though
 the above proof of the gauge invariance of ${\cal L}_{CS}$
has been given
for the case of $U_q(N)$, we point
out that all the formulas exhibited here
for {\it minimal} $U_q(N)$ continue to hold
for {\it minimal} B, C, D groups, as well as
 their inhomogeneous partners.   In particular the
commutation rule for one--forms eq.(\ref{aa})
is independent of the particular class of
algebras, when $\Lambda^2=1$, since it follows directly
from eq.(\ref{formprod}), which holds in general.
From now on whenever we  refer to a
 $q$--gauge theory, we shall assume that it is
minimally deformed, unless otherwise specified.
Nontrivial (in general multiparametric) minimal deformations
with the corresponding $q$--differential
calculus are known to exist for all groups $U(N)$ with $N>2$
and for groups of
the B, C, D type (see \cite{cast2}, \cite{sch})
together with their inhomogeneous associates.

\section{Properties of the $q$--CS term}
\setcounter{equation}{0}
Among properties 1)-3) listed at the beginning of the previous Section, 
we verified the first one for minimal deformations, using
our deformed CS Lagrangian density eq.(\ref{lag}) with $\beta=2/3$.  We
next show that the
properties 2) and 3) are also satisfied.    We begin with 3) which deals with 
the equations of motion.

 Consider a three-dimensional manifold $M$ and let $S_{CS}$
be the action obtained by integrating ${\cal L}_{CS}$ on $M$:
\be
S_{CS}=\int_M \, <dA + {2\over 3} A^2, A>  \label{csa}
\ee
In order to compute the equations of motion, we take
a variation of eq.(\ref{csa}):
\beqa
\delta S_{CS} &=& \int_M \delta <dA + {2\over 3} A^2, A> \nn \\
&=& \int_M \left(  <\delta A, dA  >  + < dA ,\delta A > \right)
+   {2\over 3} \int_M \left(  <\delta A^2, A  >  + < A^2 ,\delta A >
\right)~, \label{vact}
\eeqa
(where an integration by parts has been performed).
To proceed further we need the commutation relations for
$\delta A$ and $A$. Consider now the case of a variation
corresponding to a gauge transformation (\ref{var1}). One can
check by means of a direct computation that for a minimal deformation:
\be
\delta_{\epsilon} A^i \, A^j = -\Lambda^{ij}_{kl}\, A^k
\delta_{\epsilon} A^l~\label{cva}.
\ee
We assume that analogous commutation relations hold for
arbitrary variations of $A$ (the commutation relations for $\delta A$
with $dA$ can be obtained by taking an exterior derivative of the above
equation and assuming that the terms containing $d \delta A$ and $dA$
vanish separately).
With the help of eq.\eqn{cva}, the terms between the first pair of
parenthesis  in the right hand side
  of eq.(\ref{vact}) can be rewritten as
\be
dA^i \delta A^j (\Lambda^{kl}_{ij} <T_k, T_l> + <T_i, T_j>) ~.\label{B}
\ee
The terms between the second pair of parenthesis
in eq.(\ref{vact}) multiplied by $\frac 23$ are equal to
$$
\left({1\over 3} \delta A^i A^j A^k + {1\over 3}
A^i \delta A^j A^k \right)<[T_i, T_j],T_k>
+ {2\over3}<A^2, \delta A>=
$$
$$
\left({1\over 2} \delta A^i A^j A^k + {1\over 6}
A^i \delta A^j A^k \right)<[T_i, T_j],T_k>
+ {2\over3}<A^2, \delta A>=
$$
$$
=<\delta A, A^2> + {1\over6} A^i \delta A^j A^k 
<T_i,[T_j,T_k]>+ {2\over3}<A^2, \delta A>=
$$
\be
=<\delta A, A^2> + <A^2,\delta A>=
(A^2)^i \delta A^j (\Lambda^{kl}_{ij} <T_k, T_l> + <T_i, T_j>)~,\label{C}
\ee
where we used
$C^{k}_{ij} A^i \delta A^j = C^{k}_{ij}  \delta A^i A^j$.
Finally,
substituting eq.(\ref{B}) and eq.(\ref{C}) into eq.(\ref{vact}) we  get
\be
\delta S_{CS} = \int_M \, F^i \delta A^j \, H_{ij}
\ee
where $H_{ij}$ is the matrix
\be
H_{ij}=
\left(\Lambda^{kl}_{ij} <T_k, T_l>
+ <T_i, T_j>\right)~.
\ee
If the matrix $H_{ij}$ is nondegenerate we obtain
the desired equation of motion
\be
F=0~. \label{constraint}
\ee

Our proof of Prop. 3) thus requires 
$H_{ij}$, constructed from the $q$-group metric, and the braiding matrix
to be nondegenerate.  This condition is in addition to the minimality
assumption made earlier
on the theory.  We must therefore search for quantum
groups satisfying both of these conditions.  With regard to the
nondegeneracy requirement, we note that a nonsingular metric does not
 exist for  all classical groups; So we certainly don't expect  that
$H_{ij}$ (which reduces to twice the classical metric when $q
\rightarrow 1$)  be nondegenerate for all quantum groups.
We do  know of an example of a minimal theory with
a nondegenerate $H_{ij}$.  It has
$ISO_q(2,1)$  for the quantum gauge symmetry, and therefore
is relevant for the CS formulation
of gravity in 2+1 dimensions.  We intend  to discuss it in a future
article\cite{prep2}.

We next prove that $<F, F>$ is a closed 4--form,
\be
d <F,F>=0~, \label{4form}
\ee
and then finally  that it is exact, as stated in Prop. 2).
We use the Bianchi identity
\be
DF\equiv dF+ AF -FA=0~,\label{bianchi}
\ee
which follows from the associativity of the deformed wedge product
\cite{woro,ac}. Thus we have:
$$
d<F,F>= <DF,F>+<F,DF>-<AF-FA,F>-<F,AF-FA>=
$$
\be
=-<AF-FA,F>-<F,AF-FA>~. \label{dff}
\ee
The (ordinary) commutator  between $F$ and $A$
 can be obtained by assuming
that they are the same as in the $F=0$ case \cite{cast1}. Thus, the
$A, dA$ commutation relations are the same as the $A, A^2$ ones and,
by means of a computation completely analogous to that leading to
eq.(\ref{com3}), one can prove that:
\be
A^i F^j = \Lambda^{ij}_{kl} \, F^k A^l~.
\ee
Using the above commutators and \eqn{scalinv2} in eq. (\ref{dff}) gives:
\be
d<F,F>=F^i A^j F^k (<[T_i,T_j],T_k>-<T_i,[T_j, T_k]>)=0~.
\ee

Being closed, $<F,F>$ is locally exact. In fact it is 
equal to the exterior derivative of the CS Lagrangian density; we have:
$$
d{\cal L}_{CS}= <dA, dA> + {2\over 3} \left(<dAA- AdA, A> + <A^2, dA>
\right)=
$$
$$=<dA, dA> + \left({1\over2} dA^i A^j A^k - {1\over6}A^i dA^j A^k\right)
<[T_i,T_j], T_k> + {2\over3}<A^2, dA>=
$$
$$
=<dA, dA> + {1\over 2} dA^i A^j A^k <T_i,[T_j, T_k]>
- {1\over 6} A^i A^r dA^s \Lambda_{rs}^{jk}<[T_i,T_j], T_k>+
{2\over3}<A^2, dA>=
$$
\be
=<dA,dA> + <dA,A^2> + <A^2,dA>= <F,F>-<A^2,A^2>=<F,F>~,
\ee
since (see the Appendix)  $<A^2,A^2>=0$ for minimal deformations.

In verifying that $<F,F>=0$ is closed and exact, we have relied on
the condition of minimality. It is not clear whether or not  the
calculation can be successfully carried out for more general systems.
We know it is not possible to do for all systems.  This is because
we are able to find a counter example.  The latter corresponds to
take $U_q(2)$ as the gauge group 
which is known \cite{sch,cast1} not to  admit
nontrivial minimal deformations. We have checked, using the
explicit formulae given in \cite{cast1},
that $d<F,F> \neq 0$ for $U_q(2)$ (unless $r=q=1$, in which case there
is no deformation at all) and this implies that no CS term
can exist in this case. This appears to be a further sign
that minimality might  not only be a sufficient, but in fact, a
necessary condition for a deformed CS term to exist, although
we have not been able to prove it.

\section{Canonical Formalism}
\setcounter{equation}{0}
In this Section we present the canonical formalism for our deformed
CS action. We compute the deformed Poisson brackets among the
components of the connection and then show that the zero-curvature
constraints are the generators
of infinitesimal gauge transformations, in complete
analogy with the undeformed case. Finally, we prove that the algebra of
constraints is closed with respect to the deformed Poisson bracket.

Our procedure will be to obtain the symplectic structure starting from
the canonical one form $\theta_{{\cal L}}=\sum \delta Q P  $ of classical
mechanics.   Now since here the $Q$'s and $P$'s denote noncommuting
coordinates and momenta, care must be taken in their ordering.
(An alternative procedure would be to carry out a constraint analysis
on the noncommuting configuration space, as, for example was done in
\cite{lsy}.)

We consider the $q$--CS action (\ref{csa}) on a manifold with the
topology of a solid cylinder $M=\Sigma \times {\bf R}$,
where $\Sigma$ is some two-manifold that we think of as space
while $\bf R$ accounts for time. The action
$S_{CS}$, being a three form, is invariant under the diffeomorphisms  of $M$,
 hence it does not allow a natural choice of the time coordinate.
Since a time function is necessary in the canonical approach,
we arbitrarily choose a time function, called $x^0$, and consider
any constant $x^0$ slice (diffeomorphic to $\Sigma$)
as our space, with coordinates $\bar x \equiv (x^1,x^2)$.
According with this separation of space and time coordinates,
we split the connection $A$ in its time and space parts:
\be
A=A_0 dx^0 \,+ \, A_a dx^a \equiv A_0 dx^0 + {\bar A}~~,~~a=1, 2   \;.
\ee
(In the rest of this Section the first latin letters $a,b,\cdots$ will refer
to space coordinates.)
As it happens in the undeformed case, the $q$-CS action does not contain
time derivatives of $A_0$. We thus interpret it as a Lagrange multiplier
for the constraint:
\be
\bar F \equiv \bar d \bar A + \bar A ^2 =
{1\over 2} F_{ab} \, dx^a dx^b \approx 0~,\label{const}
\ee
where $\bar d = dx^a \partial _a$.
The phase space is then spanned by the space components of $A$, $A_1$ and
$A_2$ and we read off the canonical one--form
$\theta_{{\cal L}}$
directly from the part
of the action which contains time derivatives of these fields:
\be
\theta_{{\cal L}}=\int_{\Sigma} d^2 \bar x \,
 \delta A^i_a A^j_b \, g_{ij} \epsilon ^{ab}~~~~
\epsilon^{12}=-\epsilon^{21}=1~ 
\ee
(the fields are evaluated at the same time $x^0$).
From here we get the symplectic form
\be
\omega=\delta \theta_{{\cal L}}=
-\int_{\Sigma} d^2 \bar x \,
\delta A^i_a  \wedge \delta A^j_b \, g_{ij} \epsilon^{ab} ~,
\ee
where we apply the usual rules for exterior differentiation.
Inverting the symplectic form we get the Poisson Brackets (PB)
among the space components of the connection
\be
\{A^i_a(\bar x), A^j_b(\bar y)\}=
\epsilon_{a b}g^{ij} \delta(\bar x-\bar y)~~~~
\epsilon_{12}=-\epsilon_{21}=1, \label{PB}
\ee
$g^{ij}$ being the inverse of $g_{ij}$.
Because the deformed metric $g_{ij}$ is, in general,
non-symmetric, the PB's are, in general, not skewsymmetric.

Consider now two arbitrary function(al)s $B$ and $C$
of the fields $A_a(\bar x)$.
$B$ and $C$ are assumed to be some polynomials of the
fields $A_a(\bar x)$.  Again since $A_a(\bar x)$ are $q-$fields,
 attention must be paid to the ordering of the fields
in the definition of  $\{B,C\}$.  We set:
\be
\{B, C\} = \int \int_{\Sigma \times \Sigma} d^2\bar w d^2\bar u \,
{\delta_R B  \over \delta A^i_a (\bar w)} \,\{A^i_a(\bar w), A^j_b(\bar u)\}
\, {\delta_L C  \over \delta A^j_b (\bar u)}
\ee
where $R$ ($L$) denotes
the right (left) derivative and it is computed after pulling $A^i_a$
to the right (left).

We next investigate whether the constraints (\ref{const}) are
the generators of gauge transformations as in the undeformed case.
We first smear the field strength $\bar F$ with a Lie algebra valued function
$\epsilon=\epsilon^i T_i$
so that
\be
G (\epsilon) = \int_{\Sigma} <\epsilon, \bar F> \approx 0~.\label{gausslaw}
\ee
We assume for $\epsilon$ the commutation relations \eqn{com1}.
Then we look for the Poisson bracket of $G(\epsilon)$ with the components
of the connection
\be
\{G(\epsilon), A^j_c(\bar y)\} ~.\label{gauget}
\ee
We find
\beqa
 \{G(\epsilon), A^j_c(\bar y)\} &=&
\left\{ \int_{\Sigma} <\epsilon, \bar F>, A^j_c (\bar y) \right\}
\nn\\
&=&  \int_{\Sigma} d^2 \bar x \epsilon^k(\bar x)\left\{
(\del_a A^i_b (\bar x) +{1\over 2} C^i_{rs} A^r_a (\bar x)
A^s_b (\bar x))\, \epsilon^{a b}g_{ki}, A^j_c (\bar y) \right\}
\nn\\
&=& \del_c \epsilon^j (\bar y) - C^j_{kr} \epsilon^k(\bar y)
A_c^r(\bar y)~. \label{deltaA}
\eeqa
We can see that this is the infinitesimal transformation of the gauge
field $A^j_c (\bar y)$ with gauge parameter $-\epsilon$. Hence we have
found that $G(\epsilon)$ are the generators of the deformed gauge
transformations:
\be
\{G(\epsilon), \bar A(\bar y)\} = -\delta_{\epsilon} \bar A(\bar y)~
\label{pbga}.
\ee
As a consequence of this result the algebra of constraints \eqn{gausslaw}
closes, as can be seen by computing the PB for two constraints
(the details of the computation
can be found in the Appendix):
\be
\{G(\epsilon_1), G(\epsilon_2)\}=  G(\epsilon_1 \epsilon_2-
\epsilon_2 \epsilon_1).
\label{PBconstraints}
\ee
This result too is completely analogous to the one found in the
undeformed case.   It is consistent with the closure of gauge
transformations found in \cite{cast1} for minimal deformations.

\section{Conclusions}
As we have already stressed, the CS action we propose for minimally
deformed gauge theories has the advantage of enjoying all the topological
and algebraic features of the undeformed case. In particular,
once the existence of a new invariant scalar product has
been taken into account, a deformed Hamiltonian formalism naturally
arises, leading to a closed algebra for the appropriately smeared
constraints, which generate the corresponding gauge transformations. We
are then ready to explore the implications of the deformed CS action in
the realm of 1+1 physics by simply applying our results to a manifold
with boundary \cite{prep1}. Furthermore, as the differential calculus for
minimal $ISO_q(2,1)$ is available together with the corresponding gauge
theory, we are ready to explore the physics of $2+1$ $q$--gravity \cite{prep2}.
\vspace{.7cm}
\noindent\appendice{\Large{\bf Appendix}}

\noindent Below we derive  some equations, which hold in the case of minimal
deformations. The first one is:
\be
<A^2, A^2>=0~. \label{a2a2}
\ee
The proof consists in showing that the left hand side is proportional
to the $q$--Jacobi identity, eq.(\ref{qjac}).
\beqa
<A^2, A^2> &=& {1 \over 4 } A^i A^j A^k A^l <[T_i, T_j], [T_k, T_l]>\nn\\
&=& {1 \over 4 } A^i A^j A^k A^l <[[T_i, T_j], T_k], T_l>; \label{1st}
\eeqa
upon permuting the middle $A$'s and
using \eqn{aa}, it can be rewritten as
\be
<A^2, A^2>
= -{1 \over 4 } A^i A^j A^k A^l  \Lambda^{rs}_{jk}<[[T_i, T_r], T_s],
T_l>;
\label{2nd}
\ee
permuting the last two $A$'s it can also be rewritten as
\be
<A^2, A^2>
= {1 \over 4 } A^i A^j A^k A^l
 \Lambda^{rm}_{js} \Lambda^{sn}_{kl}<[T_i, T_r], [T_m, T_n]> \label{3rd}.
\ee
This expression can be further manipulated using \eqn{2d},
\eqn{scalinv}, \eqn{scalinv2}, in the given order. We have
\beqa
\Lambda^{rm}_{js} \Lambda^{sn}_{kl}<[T_i, T_r], [T_m, T_n]>&=&
 \Lambda^{rp}_{sl} C^s_{jk} <[T_i,T_r], T_p> \nn\\
= -<T_i, [[T_j, T_k], T_l]>&=& - <[T_i, [T_j, T_k]], T_l> ~.
\label{3rrd}
\eeqa
Putting together the three  different ways that we have found to write $<A^2,
A^2>$ , that is Eqs. \eqn{1st}, \eqn{2nd}, \eqn{3rrd}, we finally have
\be
<A^2, A^2>= {1 \over12} A^i A^j A^k A^l <\left\{ [[T_i, T_j], T_k] -
\Lambda^{rs}_{jk}[[T_i, T_r], T_s] -   [T_i, [T_j, T_k]]\right\}, T_l>
= 0
\ee
the expression between the curly brackets being zero because of 
$q$--Jacobi identity.

We now turn to the computation of the Poisson algebra of the
constraints. First, we derive the identity
\be
C_{ij}^l \Lambda^{mn}_{lk}<T_m, T_n> = <T_i, [T_j, T_k]>~.\label{newpro}
\ee
It can be proven using equations \eqn{2d}, \eqn{scalinv2}, \eqn{skewsym},
\eqn{scalinv}, \eqn{scalinv2} in the given order:
\beqa
\Lambda^{mn}_{lk} C_{ij}^l g_{mn} &=& \Lambda^{mv}_{ip} \Lambda^{pq}_{jk}
C^n_{vq} g_{mn} = \Lambda^{mv}_{ip} \Lambda^{pq}_{jk} C^n_{mv} g_{nq} \nn\\
&=& - \Lambda^{pq}_{jk} C_{ip}^m g_{mq} = C_{jk}^m g_{im}
\eeqa
which is Eq. \eqn{newpro}. Going back to the PB of two constraints,
and using eq.(\ref{pbga}), we have:
$$
\{G(\epsilon_1), G(\epsilon_2)\}=
\int_{\Sigma}
\left\{ G(\epsilon_1), <\epsilon_2(\bar y), \bar F(\bar y)> \right\}=
$$
\be
=
\int_{\Sigma}  \Lambda^{ij}_{kl}\,
\left\{ G(\epsilon_1), \bar F^k(\bar y) \right\}\, \epsilon^l_2 g_{ij}=
-\int_{\Sigma} \Lambda^{ij}_{kl}
C^k_{rs} g_{ij} \epsilon^r_1 \bar F^s
\epsilon^l_2~.
\ee
We now use eq.(\ref{newpro}) to write the above expression as:
\beqa
-\int_{\Sigma}
C^k_{sl} g_{rk} \epsilon^r_1 \bar F^s \epsilon^l_2 & = &
-\int_{\Sigma}
C^k_{sl} \Lambda^{sl}_{pq} g_{rk} \epsilon^r_1 \epsilon^p_2 \bar F^q=\cr
=\int_{\Sigma}
C^k_{pq} g_{rk} \epsilon^r_1  \epsilon^p_2 \bar F^q & = &
\int_{\Sigma}
C^k_{rp} g_{kq} \epsilon^r_1  \epsilon^p_2 \bar F^q  =\cr
& = & G(\epsilon_1 \epsilon_2- \epsilon_2 \epsilon_1)~,
\eeqa
where we have made use of eqs.(\ref{skewsym}) and (\ref{scalinv2}).

\end{document}